\documentclass{sig-alternate-05-2015}
\newcommand{\eat}[1]{}

\begin{document}

\title{Bidirectional Long-Short Term Memory\\
for Video Description}
%
%
%
%
%


\author{Yi  Bin$^1$, 
Yang  Yang$^1$, 
Zi Huang$^2$,
Fumin  Shen$^1$, 
Xing  Xu$^1$ and
Heng Tao  Shen$^{2,1}$\\
\affaddr{$^1$University of Electronic Science and Technology}\\
\affaddr{$^2$The University of Queensland}\\
\email{yi.bin@hotmail.com,~dlyyang@gmail.com,~huang@itee.uq.edu.au}\\\email{fumin.shen@gmail.com,~xing.xu@uestc.edu.cn,~shenht@itee.uq.edu.au}
} 
\maketitle
\begin{abstract}
Video captioning has been attracting broad research attention in multimedia community. However, most existing approaches either ignore temporal information among video frames or just employ local contextual temporal knowledge. In this work, we propose a novel video captioning framework, termed as \emph{Bidirectional Long-Short Term Memory} (BiLSTM), which deeply captures bidirectional global temporal structure in video. Specifically, we first devise a joint visual modelling approach to encode video data by combining a forward LSTM pass, a backward LSTM pass, together with visual features from Convolutional Neural Networks (CNNs). Then, we inject the derived video representation into the subsequent language model for initialization. The benefits are in two folds: 1) comprehensively preserving sequential and visual information; and 2) adaptively learning dense visual features and sparse semantic representations for videos and sentences, respectively. We verify the effectiveness of our proposed video captioning framework on a commonly-used benchmark, i.e., Microsoft Video Description (MSVD) corpus, and the experimental results demonstrate that the superiority of the proposed approach as compared to several state-of-the-art methods.
\end{abstract}
%
%
%
%
%
%

\section{Introduction}
With the development of digital media technology and popularity of Mobile Internet, online visual content has increased rapidly in recent couple of years. Subsequently, visual content analysis for retrieving~\cite{yang2015visual,Shen_2015_ICCV} and understanding becomes a fundamental problem in the area of multimedia research, which has motivated world-wide researchers to develop advanced techniques. Most previous works, however, have focused on classification task, such as annotating an image~\cite{krizhevsky2012imagenet,CVPR15Shen} or video~\cite{gan2015devnet,ramanathan2013video,tang2013combining,yang2014exploiting} with given fixed label sets. With some pioneering methods~\cite{farhadi2010every,ordonez2011im2text} tackling the challenge of describing images with natural language proposed, visual content understanding has attracted more and more attention. State-of-the-art techniques for image captioning have been surpassed by new advanced approaches in succession~\cite{chen2015mind,donahue2015long,karpathy2015deep,vinyals2015show,xu2015show}. Recent researches~\cite{pan2015jointly,venugopalan2014translating,yao2015describing,li2015summarization,venugopalan2015sequence} have been focusing on describing videos with more comprehensive sentences instead of simple keywords. Different from image, video is sequential data with temporal structure, which may pose significant challenge to video caption. Most of the existing works in video description employed max or mean pooling across video frames to obtain video-level representation, which failed to capture temporal knowledge. To address this problem, Yao et al. proposed to use 3-D Convolutional Neural Networks to explore local temporal information in video clips, where the most relevant temporal fragments were automatically chosen for generating natural language description with attention mechanism~\cite{yao2015describing}. In~\cite{venugopalan2015sequence}, Venugopanlan et al. implemented a Long-Short Term Memory (LSTM) network, a variant of Recurrent Neural Networks (RNNs), to model the global temporal structure in whole video snippet. However, these methods failed to exploit bidirectional
global temporal structure, which could benefit from not only previous video frames, but also information in future frames. Also, existing video captioning schemes cannot adaptively learn dense video representation and generate sparse semantic sentences.

In this work, we propose to construct a novel bidirectional LSTM (BiLSTM) network for video captioning. More specifically, we design a joint visual modelling to comprehensively explore bidirectional global temporal information in video data by integrating a forward LSTM pass, a backward LSTM pass, together with CNNs features. In order to enhance the subsequent sentence generation, the obtained visual representations are then fed into LSTM-based language model as initialization. We summarize the main contributions of this work as follows: (1) To our best knowledge, our approach is one of the first to utilize bidirectional recurrent neural networks for exploring bidirectional global temporal structure in video captioning; (2) We construct two sequential processing models for adaptive video representation learning and language description generation, respectively, rather than using the same LSTM for both video frames encoding and text decoding in~\cite{venugopalan2015sequence}; and (3) Extensive experiments on a real-world video corpus illustrate the superiority of our proposal as compared to state-of-the-arts.

\begin{figure}
\centering
\includegraphics[width=1\linewidth]{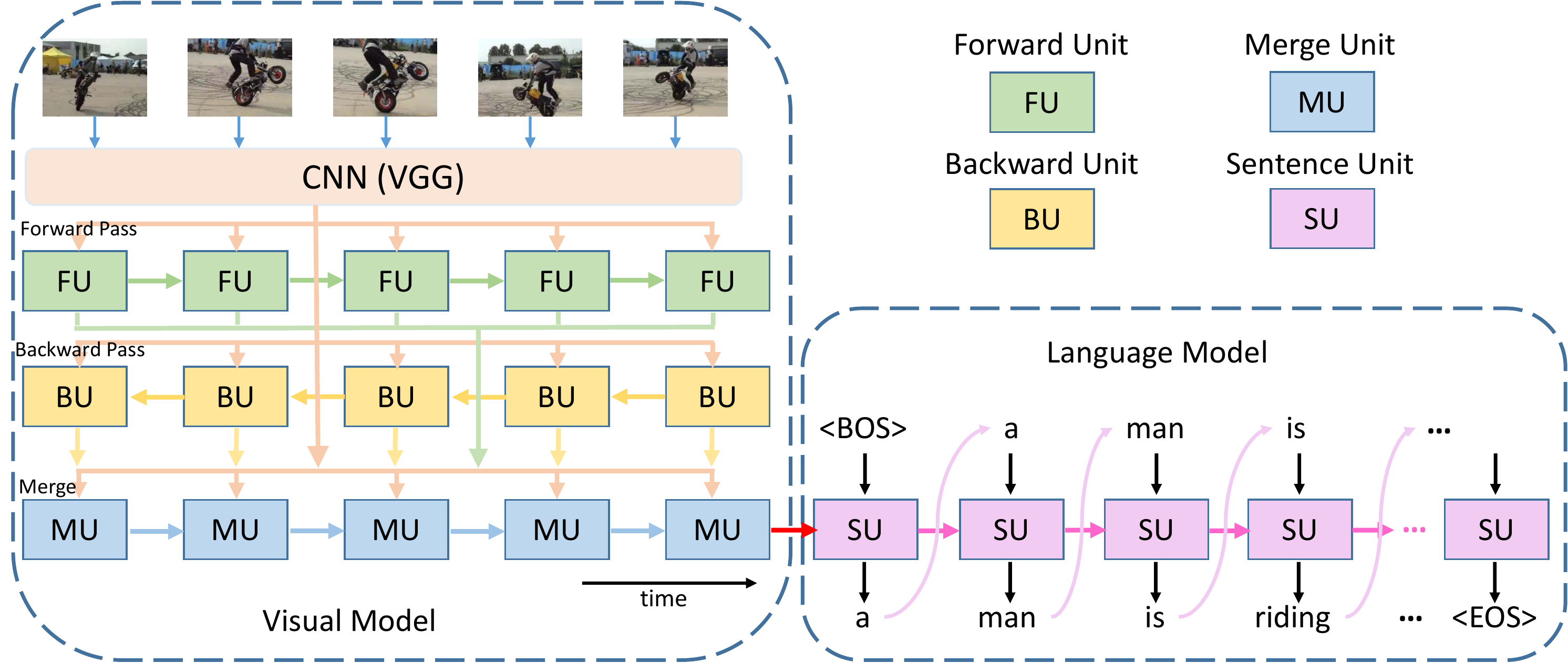}
\caption{The overall flowchart of the proposed video captioning framework. We first extract CNNs features of video frames and feed them into forward pass networks (FU, green box) and backward pass networks (BU, yellow box). We then combine the outputs of hidden states together with the original CNNs features, and pass the integrated sequence to another LSTM (MU, blue box) to generate final video representation. We initialize language model (SU, pink box) with video representation and start to generate words sequentially with <BOS> token, and terminate the process until the <EOS> token is emitted.}
\label{figure1}
\end{figure}

\section{The Proposed Approach}
In this section, we elaborate the proposed video captioning framework, including an introduction of the overall flowchart (as illustrated in Figure~\ref{figure1}), a brief review of LSTM-based Sequential Model, the joint visual modelling with bidirectional LSTM and CNNs, as well as the sentence generation process.

\subsection{LSTM-based Sequential Model}
With the success in speech recognition and machine translation tasks, recurrent neural structure, especially LSTM and its variants, have dominated sequence processing field. LSTM has been demonstrated to be able to effectively address the gradients vanishing or explosion problem~\cite{hochreiter1997long} during back-propagation through time (BPTT)~\cite{werbos1990backpropagation} and to exploit temporal dependencies in very long temporal structure. LSTM incorporates several control gates and a constant memory cell, the details of which are following:
\begin{equation}\label{1}
i_{t}=\sigma \left(W_{ix}x_{t} + W_{ih}h_{t-1}\right)
\end{equation}
\begin{equation}\label{2}
f_{t}=\sigma \left( W_{fx}x_{t} + W_{ih}h_{t-1}\right)
\end{equation}
\begin{equation}\label{3}
o_{t}=\sigma \left( W_{ox}x_{t} + W_{oh}h_{t-1}\right)
\end{equation}
\begin{equation}\label{4}
c_{t}=f_{t}\odot c_{t-1}+i_{t} \odot \phi \left ( W_{cx}x_{t}+W_{ch}h_{t-1} \right )
\end{equation}
\begin{equation}\label{5}
h_{t} = o_{t}\odot \phi \left ( c_{t} \right )
\end{equation}
where $W_{mn}$-like matrices are LSTM weight parameters, $\sigma$ and $\phi$ are denote the sigmoid and hyperbolic non-linear functions, respectively, and $\odot$ indicates element-wise multiplication operation. Inspired by the success of LSTM, we devise an LSTM-based network to investigate the video temporal structure for video representation. Then initializing language model with video representation to generate video description.
\subsection{Bidirectional Video Modelling}
Different from other video description approaches that represent video by implementing pooling across frames~\cite{venugopalan2014translating} or 3-D CNNs with local temporal structure~\cite{pan2015jointly}, we apply BiLSTM networks to exploit the bidirectional temporal structure of video clips. Convolutional Neural Networks (CNNs) has demonstrated overwhelming performance on image recognition, classification~\cite{krizhevsky2012imagenet} and video content analysis~\cite{donahue2015long,venugopalan2015sequence}. Therefore, we extract caffe~\cite{jia2014caffe} $fc7$ layer of each frame through VGG-16 layers~\cite{Simonyan14c} caffemodel. Following~\cite{venugopalan2015sequence,venugopalan2014translating}, we sample one frame from every ten frames in the video and extract the $fc7$ layer, the second fully-connected layer, to express selected frames. Then a $T$-by-4096 feature matrix generated to denote given video clip, where $T$ is the number of frames we sampled in the video. As in Figure \ref{figure1}, we then implement two LSTMs, forward pass and backward pass, to encode CNNs features of video frames, and then merge the output sequences at each time point with a learnt weight matrix. What is interesting is that at each time point in bidirectional structure, we not only ``see'' the past frames, but also ``peek'' at the future frames. In other words, our bidirectional LSTM structure encodes video by scanning the entire video sequence several times (same as the number of time steps at encoding stage), and each scan is relevant to its adjacent scans. To investigate the effect of reinforcement of original CNNs feature, we combine the merged hidden states of BiLSTM structure and $fc7$ representation time step-wise. We further employ another forward pass LSTM network with incorporated sequence to generate our video representation. In~\cite{wu2015fusing,wu2015modeling}, Wu et al. had demonstrated that using the output of the last step could perform better than pooling approach across outputs of all the time steps in video classification task. Similarly, we represent the entire video clip using the state of memory cell and output of the last time point, and feed them into description generator as initialization of memory cell and hidden unit respectively.
\subsection{Generating Video Description}
Existing video captioning approaches usually share common part of visual model and language model as representation~\cite{venugopalan2015sequence,pan2015jointly}, which may lead to severe information loss. Besides, they also input the same pooled visual vector of the whole video into every sentence processing unit, thereby ignoring temporal structure. Such methods may easily result in undesirable outputs due to the duplicate inputs in every time point of the new sequence~\cite{venugopalan2014translating}. To address these issues, we generate descriptions for video clips using a sequential model initialized with visual representation. Inspired by the superior performance of probabilistic sequence generation machine, we generate each word recurrently at each time point. Then the log probability of sentence $S$ can be expressed as below:
\begin{equation}\label{7}
\log p\left ( S|V \right )=\sum_{t=1}^{t=N}\log p\left ( w_{t}|V,w_{1},...w_{t-1};\theta \right )
\end{equation}
where $\theta$ denotes all parameters in sentence generation model and $V$ is the representation of given video, and $N$ indicates the number of words in sentence. We identify the most likely sentence by maximizing the log likelihood in Equation (\ref{7}), then our object function can be described as:
\begin{equation}\label{8}
\theta^* = \underset{\theta}{\text{argmax\ }}\sum_{t=1}^{t=N}\log p\left ( S|V;\theta \right )
\end{equation}
The optimizer updates $\theta$ with $\theta^*$ across the entire training process applying Stochastic Gradient Descent (SGD). During training phrase, the loss is back propagated through time and each LSTM unit learns to derive an appropriate hidden representation $h_{t}$ from input sequence. We then implement the Softmax function to get the probability distribution over the words in the entire vocabulary.

At the beginning of the sentence generation, as depicted in Figure \ref{figure1}, an explicit starting token (<BOS>) is needed and we terminate each sentence when the end-of-sentence token (<EOS>) is feeding in. During test phrase, similar to~\cite{venugopalan2015sequence}, our language model takes the word $w_{t-1}$ with maximum likelihood as input at time $t$ repeatedly until the <EOS> token is emitted.

\section{Experiments}
\subsection{Dataset}
\textbf{Video Dataset}: We evaluate our approach by conducting experiments on the Microsoft Research Video Description (MSVD)~\cite{chen2011collecting} corpus, which is description for a collection of 1,970 video clips. Each video clip depicts a single action or a simple event, such as ``shooting'', ``cutting'', ``playing the piano'' and ``cooking'', which with the duration between 8 seconds to 25 seconds. There are roughly 43 available sentences per video and 7 words in each sentence at average. Following the majority of prior works~\cite{pan2015jointly,venugopalan2014translating,venugopalan2015sequence,yao2015describing}, we split entire dataset into training, validation and test set with 1200, 100 and 670 snippets, respectively.

\textbf{Image Dataset}: Comparing to other LSTM structure and deep networks, the size of video dataset for caption task is small, thereby we apply transferring learning from image description. COCO 2014 image description dataset~\cite{lin2014microsoft} has been used to perform experiments frequently~\cite{karpathy2015deep,donahue2015long,chen2015mind,xu2015show}, which consists of more than 120,000 images, about 82,000 and 40,000 images for training and test respectively. We pre-train our language model on COCO 2014 training set first, then transfer learning on MSVD with integral video description model.

\subsection{Experimental Setup}
\subsubsection{Preprocessing}

\textbf{Description Processing}: Some minimal preprocessing has been implemented to the descriptions in both MSVD and COCO 2014 datasets. We first employ \textsl{word\textunderscore tokenize} operation in NLTK toolbox\footnote{\url{http://www.nltk.org}} to obtain individual words, and then convert all words to lower-case. All punctuation are removed, and then we start each sentence with <BOS> and end with <EOS>. Finally, we combine the sets of words in MSVD with COCO 2014, and generate a vocabulary with 12,984 unique words. Each word input to our system is represented by one-hot vector.

\textbf{Video Preprocessing}:
As previous video description works~\cite{venugopalan2014translating,venugopalan2015sequence,pan2015jointly} , we sample video frames once in every ten frames, then these frames could represent given video and 28.5 frames for each video averagely. We extract frame-wise caffe $fc7$ layer features using VGG-16 layers model, then feed the sequential feature into our video caption system.

\subsubsection{Model}
We employ a bidirectional S2VT~\cite{venugopalan2015sequence} and a joint bidirectional LSTM structure to investigate the performance of our bidirectional approach. For convenient comparison, we set the size of hidden unit of all LSTMs in our system to 512 as~\cite{pan2015jointly,venugopalan2015sequence}, except for the first video encoder in unidirectional joint LSTM. During training phrase, we set 80 as maximum number of time steps of LSTM in all our models and a mini-batch with 16 video-sentence pairs. We note that over 99\% of the descriptions in MSVD and COCO 2014 contain no more than 40 words, and in~\cite{venugopalan2015sequence}, Venugopalan et al. pointed out that 94\% of the YouTube training videos satisfy our maximum length limit. To ensure sufficient visual content, we adopt two ways to truncate the videos and sentences adaptively when the sum of the number of frames and words exceed the limit. If the number of words is within 40, we arbitrarily truncate the frames to satisfy the maximum length. When the length of sentence is more than 40, we discard the words that beyond the length and take video frames with a maximum number of 40.

\textbf{Bidirectional S2VT}: Similar to~\cite{venugopalan2015sequence}, we implement several S2VT-based models: S2VT, bidirectional S2VT and reinforced S2VT with bidirectional LSTM video encoder. We conduct experiment on S2VT using our video features and LSTM structure instead of the end-to-end model in~\cite{venugopalan2015sequence}, which need original RGB frames as input. For bidirectional S2VT model, we first pre-train description generator on COCO 2014 for image caption. We next implement forward and backward pass for video encoding and merge the hidden states step-wise with a learnt weight while the language layer receives merged hidden representation with null padded as words. We also pad the inputs of forward LSTM and backward LSTM with zeros at decoding stage, and concatenate the merged hidden states to embedded words. In the last model, we regard merged bidirectional hidden states as complementary enhancement and concatenate to original $fc7$ features to obtain a reinforced representation of video, then derive sentence from new feature using the last LSTM. The loss is computed only at decoding stage in all S2VT-based models.

\textbf{Joint-BiLSTM}: Different from S2VT-based models, we employ a joint bidirectional LSTM networks to encode video sequence and decode description applying another LSTM respectively rather than sharing the common one. We stack two layers of LSTM networks to encode video and pre-train language model as in S2VT-based models. Similarly, unidirectional LSTM, bidirectional LSTM and reinforced BiLSTM are executed to investigate the performance of each structure. We set 1024 hidden units of the first LSTM in unidirectional encoder so that the output could pass to the second encoder directly, and the memory cell and hidden state of the last time point are applied to initialize description decoder. Bidirectional structure and reinforced BiLSTM in encoder are implemented similarly to the corresponding type structure in S2VT-based models, respectively, and then feed the video representation into description generator as the unidirectional model aforementioned.
\begin{figure}
\centering
\includegraphics[width=1\linewidth]{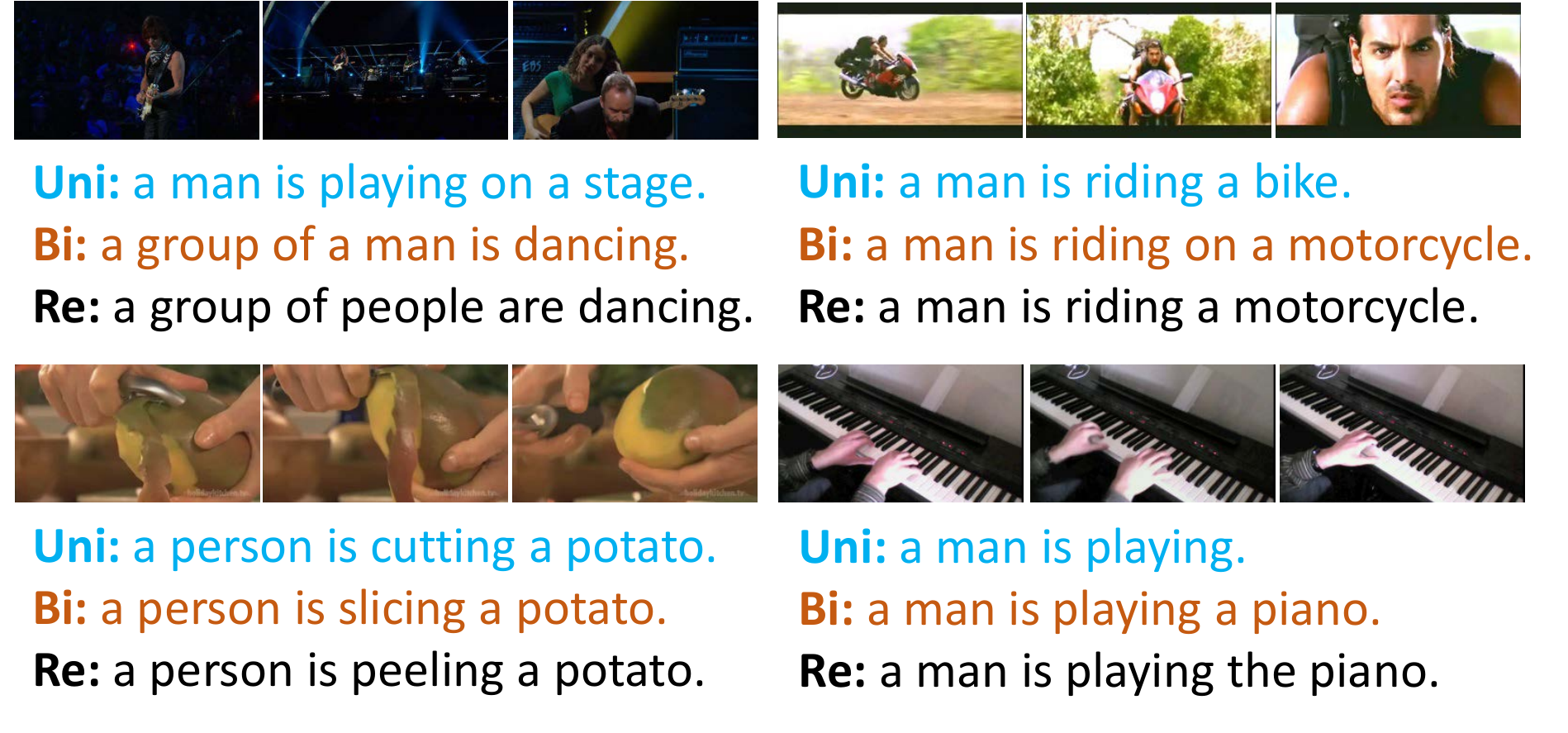}
\caption{Video captioning examples of our proposed method. ``Uni'' in color blue, ``Bi'' in color brown and ``Re'' in color black are unidirectional Joint-LSTM, bidirectional Joint-LSTM and reinforced Joint-BiLSTM model, respectively.}\label{figure2}
\end{figure}
\subsection{Results and Analysis}
BLEU~\cite{papineni2002bleu}, METEOR~\cite{lavie2014meteor}, ROUGE-L~\cite{lin2004rouge} and CIDEr~\cite{vedantam2015cider} are common evaluation metrics in image and video description, the first three were originally proposed to evaluate machine translation at the earliest and CIDEr was proposed to evaluate image description with sufficient reference sentences. To quantitatively evaluate the performance of our bidirectional recurrent based approach, we adopt METEOR metric because of its robust performance. Contrasting to the other three metrics, METEOR could capture semantic aspect since it identifies all possible matches by extracting exact matcher, stem matcher, paraphrase matcher and synonym matcher using WordNet database, and compute sentence level similarity scores according to matcher weights. The authors of CIDEr also argued for that METEOR outperforms CIDEr when the reference set is small~\cite{vedantam2015cider}.

We first compare our unidirectional, bidirectional structures and reinforced BiLSTM. As shown in Table \ref{table1}, in S2VT-based model, bidirectional structure performs very little lower score than unidirectional structure while it shows the opposite results in joint LSTM case. It may be caused by the pad at description generating stage in S2VT-based structure. We note that BiLSTM reinforced structure gains more than 3\% improvement than unidirectional-only model in both S2VT-based and joint LSTMs structures, which means that combining bidirectional encoding of video representation is beneficial to exploit some additional temporal structure in video encoder (Figure \ref{figure2}). On structure level, Table \ref{table1} illustrates that our Joint-LSTMs based models outperform all S2VT based models correspondingly. It demonstrates our Joint-LSTMs structure benefits from encoding video and decoding natural language separately.

\begin{table}
\centering
\caption{Comparison results of unidirectional, bidirectional structures and reinforced BiLSTM in both S2VT-based and joint LSTMs structure with METEOR (reported in percentage, higher is better).}
\begin{tabular}{l|c} \hline
\textbf{Model}&\textbf{METEOR}\\ \hline
S2VT-unidirectional & 28.7\\
S2VT-bidirectional & 28.6\\
S2VT-BiLSTM reinfored & \textbf{29.5}\\ \hline
Joint-LSTM unidirectional & 29.5\\
Joint-LSTM bidirectional & 29.9\\
Joint-BiLSTM reinforced & \textbf{30.3}\\
\hline\end{tabular}
\label{table1}
\end{table}

We also evaluate our Joint-BiLSTM structure by comparing with several other state-of-the-art baseline approaches, which exploit either local or global temporal structure. As shown in Table \ref{table2}, our Joint-BiLSTM reinforced model outperforms all of the baseline methods. The result of ``LSTM'' in first row refer from~\cite{pan2015jointly} and the last row but one denotes the best model combining local temporal structure using C3D with global temporal structure utilizing temporal attention in~\cite{yao2015describing}. From the first two rows, our unidirectional joint LSTM shows rapid improvement, and comparing with S2VT-VGG model in line 3, it also demonstrates some superiority. Even LSTM-E jointly models video and descriptions representation by minimizing the distance between video and corresponding sentence, our Joint-BiLSTM reinforced obtains better performance from bidirectional encoding and separated visual and language models.
\begin{table}
\centering
\caption{Comparing with several state-of-the-art models (reported in percentage, higher is better).}
\begin{tabular}{l|c} \hline
\textbf{Model}&\textbf{METEOR}\\ \hline
LSTM & 26.9\\
Joint-LSTM unidirectinal (ours)& 29.5\\ \hline
S2VT~\cite{venugopalan2015sequence}\\
-RGB (VGG)& 29.2\\
-RGB (VGG)+Flow (AlexNet) & 29.8\\ \hline
LSTM-E (VGG)~\cite{pan2015jointly} & 29.5\\
LSTM-E (C3D)~\cite{pan2015jointly} & 29.9\\ \hline
Yao et al.~\cite{yao2015describing} & 29.6\\ \hline
Joint-BiLSTM reinforced (ours) & \textbf{30.3}\\
\hline\end{tabular}
\label{table2}
\end{table}

We observed that while our unidirectional S2VT has the same deployment as~\cite{venugopalan2015sequence}, our model gives a little poorer performance(line 1, Table~\ref{table1} and line 3, Table~\ref{table2}). As mentioned in Section 3.2.2, they employed an end-to-end model reading original RGB frames and fine-tuning on the VGG caffemodel. The features of frames from VGG $fc7$ layer are more compatible to MSVD dataset and the description task. However, our joint LSTM demonstrates better performance with general features rather than specific ones for data, even superior to their model with multiple feature aspects (RGB + Flow, line 4, Table \ref{table2}), which means that our Joint-BiLSTM could show more powerful descriptive ability in end-to-end case. Certainly, We would investigate effect of end-to-end type of our Joint-BiLSTM in future works.

\section{Conclusion and Future Works}
In this paper, we introduced a sequence to sequence approach to describe video clips with natural language. The core of our method was, we applied two LSTM networks for the visual encoder and natural language generator component of our model. In particular, we encoded video sequences with a bidirectional Long-Short Term Memory (BiLSTM) network, which could effectively capture the bidirectional global temporal structure in video. Experimental results on MSVD dataset demonstrated the superior performance over many other state-of-the-art methods.

We also note some limitations in our model, such as end-to-end framework employed in~\cite{venugopalan2015sequence} and distance measured in~\cite{pan2015jointly}. In the future we will make more effort to fix these limitations and exploit the linguistic domain knowledge in visual content understanding.

\bibliographystyle{abbrv}
\bibliography{2016mm}

\begin{thebibliography}{10}

\bibitem{chen2011collecting}
D.~L. Chen and W.~B. Dolan.
\newblock Collecting highly parallel data for paraphrase evaluation.
\newblock In {\em ACL}, 2011.

\bibitem{chen2015mind}
X.~Chen and C.~Lawrence~Zitnick.
\newblock Mind's eye: A recurrent visual representation for image caption
  generation.
\newblock In {\em CVPR}, 2015.

\bibitem{donahue2015long}
J.~Donahue, L.~Anne~Hendricks, S.~Guadarrama, M.~Rohrbach, S.~Venugopalan,
  K.~Saenko, and T.~Darrell.
\newblock Long-term recurrent convolutional networks for visual recognition and
  description.
\newblock In {\em CVPR}, 2015.

\bibitem{farhadi2010every}
A.~Farhadi, M.~Hejrati, M.~A. Sadeghi, P.~Young, C.~Rashtchian, J.~Hockenmaier,
  and D.~Forsyth.
\newblock Every picture tells a story: Generating sentences from images.
\newblock In {\em ECCV}. Springer, 2010.

\bibitem{gan2015devnet}
C.~Gan, N.~Wang, Y.~Yang, D.-Y. Yeung, and A.~G. Hauptmann.
\newblock Devnet: A deep event network for multimedia event detection and
  evidence recounting.
\newblock In {\em CVPR}, 2015.

\bibitem{hochreiter1997long}
S.~Hochreiter and J.~Schmidhuber.
\newblock Long short-term memory.
\newblock {\em Neural computation}, 9(8):1735--1780, 1997.

\bibitem{jia2014caffe}
Y.~Jia, E.~Shelhamer, J.~Donahue, S.~Karayev, J.~Long, R.~Girshick,
  S.~Guadarrama, and T.~Darrell.
\newblock Caffe: Convolutional architecture for fast feature embedding.
\newblock In {\em ACM Multimedia}, 2014.

\bibitem{karpathy2015deep}
A.~Karpathy and L.~Fei-Fei.
\newblock Deep visual-semantic alignments for generating image descriptions.
\newblock In {\em CVPR}, 2015.

\bibitem{krizhevsky2012imagenet}
A.~Krizhevsky, I.~Sutskever, and G.~E. Hinton.
\newblock Imagenet classification with deep convolutional neural networks.
\newblock In {\em NIPS}, 2012.

\bibitem{lavie2014meteor}
M.~D.~A. Lavie.
\newblock Meteor universal: language specific translation evaluation for any
  target language.
\newblock {\em ACL}, 2014.

\bibitem{li2015summarization}
G.~Li, S.~Ma, and Y.~Han.
\newblock Summarization-based video caption via deep neural networks.
\newblock In {\em ACM Multimedia}, 2015.

\bibitem{lin2004rouge}
C.-Y. Lin.
\newblock Rouge: A package for automatic evaluation of summaries.
\newblock In {\em ACL}, 2004.

\bibitem{lin2014microsoft}
T.-Y. Lin, M.~Maire, S.~Belongie, J.~Hays, P.~Perona, D.~Ramanan,
  P.~Doll{\'a}r, and C.~L. Zitnick.
\newblock Microsoft coco: Common objects in context.
\newblock In {\em ECCV}. Springer, 2014.

\bibitem{ordonez2011im2text}
V.~Ordonez, G.~Kulkarni, and T.~L. Berg.
\newblock Im2text: Describing images using 1 million captioned photographs.
\newblock In {\em NIPS}, 2011.

\bibitem{pan2015jointly}
Y.~Pan, T.~Mei, T.~Yao, H.~Li, and Y.~Rui.
\newblock Jointly modeling embedding and translation to bridge video and
  language.
\newblock {\em arXiv preprint arXiv:1505.01861}, 2015.

\bibitem{papineni2002bleu}
K.~Papineni, S.~Roukos, T.~Ward, and W.-J. Zhu.
\newblock Bleu: a method for automatic evaluation of machine translation.
\newblock In {\em ACL}, 2002.

\bibitem{ramanathan2013video}
V.~Ramanathan, P.~Liang, and L.~Fei-Fei.
\newblock Video event understanding using natural language descriptions.
\newblock In {\em ICCV}, 2013.

\bibitem{Shen_2015_ICCV}
F.~Shen, W.~Liu, S.~Zhang, Y.~Yang, and H.~Tao~Shen.
\newblock Learning binary codes for maximum inner product search.
\newblock In {\em ICCV}, pages 4148--4156, December 2015.

\bibitem{CVPR15Shen}
F.~Shen, C.~Shen, W.~Liu, and H.~T. Shen.
\newblock Supervised discrete hashing.
\newblock In {\em CVPR}, pages 37--45, 2015.

\bibitem{Simonyan14c}
K.~Simonyan and A.~Zisserman.
\newblock Very deep convolutional networks for large-scale image recognition.
\newblock {\em CoRR}, abs/1409.1556, 2014.

\bibitem{tang2013combining}
K.~Tang, B.~Yao, L.~Fei-Fei, and D.~Koller.
\newblock Combining the right features for complex event recognition.
\newblock In {\em ICCV}, 2013.

\bibitem{vedantam2015cider}
R.~Vedantam, C.~Lawrence~Zitnick, and D.~Parikh.
\newblock Cider: Consensus-based image description evaluation.
\newblock In {\em CVPR}, 2015.

\bibitem{venugopalan2015sequence}
S.~Venugopalan, M.~Rohrbach, J.~Donahue, R.~Mooney, T.~Darrell, and K.~Saenko.
\newblock Sequence to sequence-video to text.
\newblock In {\em ICCV}, 2015.

\bibitem{venugopalan2014translating}
S.~Venugopalan, H.~Xu, J.~Donahue, M.~Rohrbach, R.~Mooney, and K.~Saenko.
\newblock Translating videos to natural language using deep recurrent neural
  networks.
\newblock {\em arXiv preprint arXiv:1412.4729}, 2014.

\bibitem{vinyals2015show}
O.~Vinyals, A.~Toshev, S.~Bengio, and D.~Erhan.
\newblock Show and tell: A neural image caption generator.
\newblock In {\em CVPR}, 2015.

\bibitem{werbos1990backpropagation}
P.~J. Werbos.
\newblock Backpropagation through time: what it does and how to do it.
\newblock {\em Proceedings of the IEEE}, 78(10):1550--1560, 1990.

\bibitem{wu2015fusing}
Z.~Wu, Y.-G. Jiang, X.~Wang, H.~Ye, X.~Xue, and J.~Wang.
\newblock Fusing multi-stream deep networks for video classification.
\newblock {\em arXiv preprint arXiv:1509.06086}, 2015.

\bibitem{wu2015modeling}
Z.~Wu, X.~Wang, Y.-G. Jiang, H.~Ye, and X.~Xue.
\newblock Modeling spatial-temporal clues in a hybrid deep learning framework
  for video classification.
\newblock In {\em ACM Multimedia}, 2015.

\bibitem{xu2015show}
K.~Xu, J.~Ba, R.~Kiros, A.~Courville, R.~Salakhutdinov, R.~Zemel, and
  Y.~Bengio.
\newblock Show, attend and tell: Neural image caption generation with visual
  attention.
\newblock {\em arXiv preprint arXiv:1502.03044}, 2015.

\bibitem{yang2014exploiting}
Y.~Yang, Z.-J. Zha, Y.~Gao, X.~Zhu, and T.-S. Chua.
\newblock Exploiting web images for semantic video indexing via robust
  sample-specific loss.
\newblock {\em TMM}, 16(6):1677--1689, 2014.

\bibitem{yang2015visual}
Y.~Yang, H.~Zhang, M.~Zhang, F.~Shen, and X.~Li.
\newblock Visual coding in a semantic hierarchy.
\newblock In {\em ACM Multimedia}, pages 59--68, 2015.

\bibitem{yao2015describing}
L.~Yao, A.~Torabi, K.~Cho, N.~Ballas, C.~Pal, H.~Larochelle, and A.~Courville.
\newblock Describing videos by exploiting temporal structure.
\newblock In {\em ICCV}, 2015.

\end{thebibliography}
\end{document}